\documentclass[%
 reprint,
superscriptaddress,
 amsmath,amssymb,
 aps,
floatfix,
]{revtex4-2}
\usepackage{float}
\usepackage{graphicx}
\usepackage{dcolumn}
\usepackage{bm}

\usepackage[utf8]{inputenc}
\usepackage[T1]{fontenc}
\usepackage{mathptmx}
\usepackage{etoolbox}
\usepackage{xcolor}
\usepackage{soul}
\usepackage[squaren,Gray]{SIunits}

\begin{document}
\preprint{APS}
\title[]{
 Acoustic scattering singularities via quasi-Bound states in the continuum
}

\author{Anis Maddi}
\affiliation{Université de Lorraine, CNRS, IJL, F-54000 Nancy, France}
\email{anis.maddi@univ-lorraine.fr}
 \author{Mourad Oudich}
 \affiliation{Université de Lorraine, CNRS, IJL, F-54000 Nancy, France}
  \affiliation{Institut Universitaire de France (IUF), F-75231, Paris, France}
   \author{Aurelien Merkel }
 \author{Julio A. Iglesias Martínez}

 \author{Badreddine Assouar}
 \email{badreddine.assouar@univ-lorraine.fr}

\affiliation{Université de Lorraine, CNRS, IJL, F-54000 Nancy, France}

\date{\today} 

\begin{abstract}

Non-Hermitian systems enable advanced control of wave propagation by exploiting engineered losses. This introduces an additional degree of freedom that permits the emergence of exceptional points (EPs). In this letter, we theoretically and experimentally demonstrate the control of scattering singularities in a non-Hermitian acoustic system using quasi–bound states in the continuum (qBICs). Through Friedrich–Wintgen interference, the losses of a two-port cavity are tuned until achieving critical coupling, yielding narrowband coherent perfect absorption (CPA) with a quality factor of $140$. Additionally, by coupling two distinct resonators, we observe the emergence of an EP, where both eigenvalues simultaneously coalesce and vanish, resulting in narrowband unidirectional absorption. Our results establish a connection between qBICs and scattering singularities, and offer a route toward acoustic devices featuring narrowband resonances and tunable radiative losses.

\end{abstract}

\maketitle

\newpage

\textit{Introduction.—}  In recent years, interest in the compelling features of non-Hermitian Hamiltonians has grown substantially.
Non-Hermitian systems \cite{ashida2020non,el2018non} describe non-conservative dynamics induced by the presence of gain and loss elements. This surge of interest stems from early work on parity–time (PT) symmetric systems, which demonstrated that a careful balance between gain and loss enables unconventional phenomena, such as invisible sensing \cite{fleury2015invisible} and coherent perfect absorption–lasing (CPA-lasing) \cite{auregan2017pt,poignand2021parity}.
Inspired by these developments, studies of non-Hermitian systems have expanded to explore novel physics arising from exceptional points (EPs) \cite{wang2019extremely,achilleos2017non}, where eigenvalues and eigenstates coalesce, as well as robust wave transport induced by non-Hermitian topology \cite{ding2022non,zhang2021acoustic,maddi2024exact}.

Tailoring losses is essential for designing structures that manipulate waves \cite{jin20252024}. In acoustics, minimizing these losses remains challenging \cite{cummer2016controlling} and may constrain experimental implementations of novel phenomena. 
In conventional systems, acoustic resonances typically exhibit short lifetimes due to radiative and viscothermal dissipation. A promising route to overcome this limitation is provided by bound states in the continuum (BICs), which achieve high quality factors ($Q$ factors) by strongly suppressing radiation leakage \cite{hsu2016bound}. BICs are modes that exist within the continuum of propagating modes but remain completely decoupled from the radiation channels, resulting in localized modes. They can emerge through several mechanisms \cite{huang2022general,an2024multibranch,kronowetter2023realistic}, such as Friedrich–Wintgen (FW) and Fabry–Pérot (FP)  interferences. When slightly disturbed, BICs become  weakly coupled to the radiation channels, thus turning into quasi-bound states in the continuum (qBICs). Resonant structures supporting qBICs have been explored across multiple physical platforms, including optical \cite{hsu2013observation,plotnik2011experimental,pankin2020one,koshelev2019meta}, acoustic \cite{marti2024observation,farhat2024observation,chen2023phonon,kronowetter2025exceptional,huang2020extreme,krasikova2024acoustic,zhang2023topological}, and elastic systems \cite{lee2023elastic,cao2021elastic,marti2023bound}. Although qBICs have been employed in acoustics to realize several resonance-based phenomena, their implementation for airborne absorption problems has received limited attention, particularly in multi-port systems.

CPA is an interesting strategy for waves absorption, as it enables the complete suppression of outgoing waves in multi-port systems. CPA, also known as anti-lasing,  corresponds to a time-reversed spectral singularity,
occurring when the zero of the scattering matrix crosses the real frequency axis. 
This condition is satisfied when the intrinsic $\gamma_{int}$ and radiative $\gamma_{rad}$ decay rates are perfectly balanced $\gamma_{rad}=\gamma_{int}$, a condition known as critical coupling \cite{maddi2025direct,huang2024acoustic,romero2016perfect}. CPA implementations in acoustics \cite{farooqui2022ultrathin,baranov2017coherent,olivier2022asymmetric,achilleos2016coherent,merkel2015control} typically rely on resonators with low  $Q$ factors to match the strong radiative leakage, which limits their potential for strong sound confinement or their implementation as narrowband filters.

\begin{figure}
    \centering
    \includegraphics[width=1\linewidth]{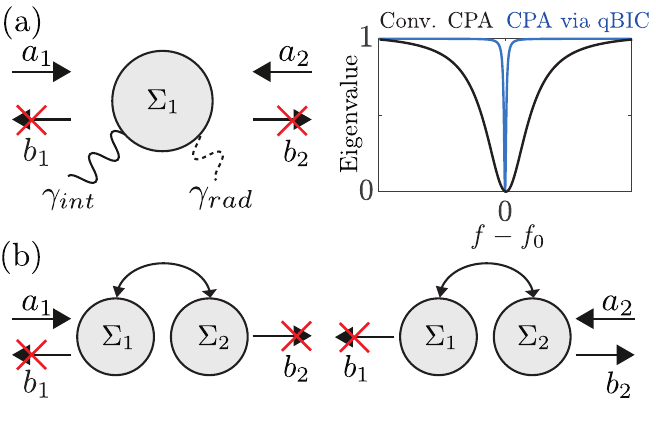}
    \caption{Interplay between qBICs and absorption in two-port scattering problems with input $(a_1,a_2)$ and output $(b_1,b_2)$ waves.  (a) Resonator $\Sigma_1$ with resonance frequency $f_0$, radiative $\gamma_{rad}$ and intrinsic $\gamma_{int}$ decay rates.  A conventional resonator under critical coupling conditions yields CPA when $\gamma_{int}=\gamma_{rad}$ but typically with a moderate or low $Q$ factor. Once a qBIC is employed, the critical coupling condition can be satisfied with significantly lower $\gamma_{rad}$, resulting in narrowband absorption. 
    (b) EP combining qBIC resonance and absorption using a system of coupled resonators $\Sigma_1$ and $\Sigma_2$. In this configuration, the eigenvalues and eigenvectors simultaneously vanishes, resulting in a unidirectional absorber. Waves impinging from the left are fully absorbed while waves propagating from the right are fully reflected.}
    \label{Fig1}
    \end{figure}

In this letter, we extend the framework of acoustic qBICs by demonstrating their potential to exploit scattering singularities in the form of CPA and an EP. We show that the combination of a qBIC and CPA in an acoustic resonator yields narrowband perfect absorption, as illustrated in Fig.~\ref{Fig1}.(a). The critical coupling condition is achieved by fine-tuning the resonator geometry, which controls FW interference and the radiative decay rate. Experimentally, CPA via qBIC is implemented in a rectangular cavity with two radiation channels, yielding a sharp resonance with a $Q$ factor of $140$. Afterward, we demonstrate the appearance of an EP using two distinct resonators (see Fig.~\ref{Fig1}.(b)). In this configuration, the eigenvalues of the scattering matrix coalesce and simultaneously vanish, thereby combining qBIC resonance with absorption at an EP. Experimentally, the setup consists of two resonators coupled through an intermediate channel. At a specific channel length, FP interference yields an EP, resulting in a narrowband unidirectional absorption of 99\%.

\begin{figure*}
    \centering
    \includegraphics[width=1\linewidth]{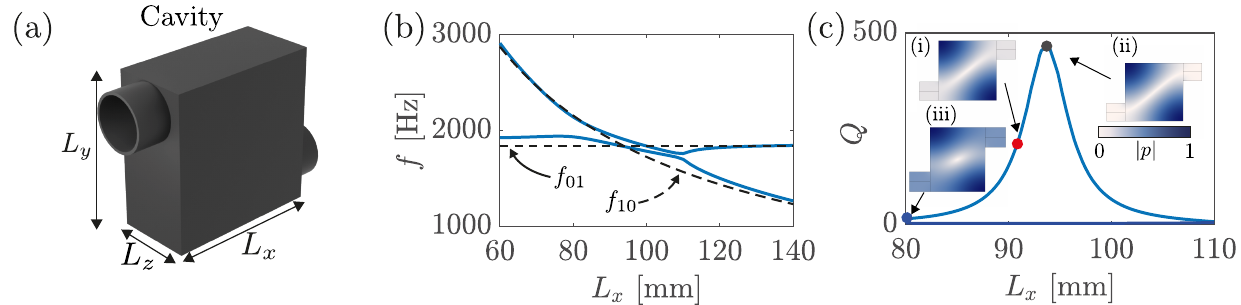}
    \caption{Formation of a qBIC in an acoustic cavity. (a) Cavity of dimension $L_x$, $L_y$ and $L_z$ with two radiation channels. (b) Resonance frequencies 
    for varying cavity length $L_x$. The dashed lines represents the resonance frequencies $f_{01}$ and $f_{10}$ in the absence of coupling. (c) $Q$ factors for varying $L_x$. Insets (i--iii) represents the pressure field magnitude  (mode shapes) for $L_x = 93.6\,\text{mm}$, $L_x = 91\,\text{mm}$, and $L_x = 80\,\text{mm}$.   }
    \label{Figure2}
\end{figure*}

\textit{Loss control mechanism.—}  
To start, we consider the interaction between two modes in a single open cavity operating above the first cut-off frequency. This mode coupling, known as FW interference, originates from mode interferences near an avoided degeneracy and can be engineered by tuning the system's resonances, for instance, through geometric modifications of the cavity~\cite{kronowetter2023realistic,huang2022general}. In acoustics, such FW interference has been reported to arise from the coupling between rectangular cavity modes. Building on this concept, we investigate a rectangular cavity of dimensions $L_x$, $L_y$, and $L_z$ connected to two radiation channels, as illustrated in Fig.~\ref{Figure2}(a). For simplicity, we restrict our analysis to the interaction between the first two modes, $\psi_{10}$ and $\psi_{01}$, with resonance frequencies $f_{10} = c/(2L_x)$ and $f_{01} = c/(2L_y)$, respectively, and focus on operating frequencies below the cut-off of the radiation channels.

The interaction between the first two acoustic modes is characterized through an eigenfrequency analysis for varying cavity length $L_x$, with $L_y = 94\text{mm}$ and $L_z=44\text{mm}$, as shown in Fig.~\ref{Figure2}(b,c). The numerical simulations are performed using the commercial software COMSOL Multiphysics. As expected from a FW resonance, the evolution of the resonance frequencies with respect to a parametric change ($L_x$) exhibits an avoided crossing close to the degeneracy point $L_x \approx L_y$, as shown in Fig.~\ref{Figure2}(b). This interaction strongly alters the radiative properties of the system, giving rise to a hybridized mode with suppressed radiation losses. Consequently, a narrowband resonance emerges, reaching a maximum $Q$ factor of $Q_{\text{max}} = 467$ at the optimal cavity length $L_x = 93.6\,\text{mm}$ (Fig.~\ref{Figure2}(c)).

The pronounced suppression of radiation losses leads to strong confinement of acoustic energy inside the cavity, preventing it from leaking into the surrounding channels. For instance, the mode shape (see Fig. 2(d)) corresponds to the configuration resulting in the highest $Q$ factor and exhibits strong confinement within the cavity. The second configuration (see Fig 2(e)) corresponds to the mode for $L_x=91$ mm, and displays similar features, namely a high $Q$ factor and localization, albeit with a smaller $Q$ factor due to slight energy leakage. In contrast, the mode at $ L_x= 80$mm (see Fig. 2(f)) radiates significantly more into the channels, resulting in a much lower $Q$ factor.

Therefore, by tuning the geometrical parameters, one can control the radiation losses. This property is particularly appealing, as it provides a means to precisely manipulate wave propagation. Such control can be exploited in applications like wave absorption, where accurate tuning of losses is essential.

\textit{CPA using qBIC.—}  
Let us consider  a reciprocal and mirror symmetric two-port acoustic system that can be described by a scattering matrix  $\mathbf{S}$. This matrix relates the outgoing waves $v_{out}$ to the incoming waves $v_{in}$ as follows,

\begin{equation}
    \underbrace{\begin{pmatrix}
b_2  \\
b_1
\end{pmatrix}}_{v_{out}}=\underbrace{\begin{pmatrix}
t & r \\
r & t
\end{pmatrix}}_{\mathbf{S}} \underbrace{\begin{pmatrix}
a_1  \\
a_2 
\end{pmatrix} }_{v_{in}},
\end{equation}

where $t$ and $r$ are the transmission and reflection coefficients, respectively. The matrix has two eigenvalues $\lambda_{\pm}=r\pm t$ associated with the eigenvectors $v_{\pm}=[1,\pm 1]^T$.

In order to completely suppress the outgoing waves $v_{out}$, the scattering matrix must satisfy the following eigenvalue problem,

\begin{equation}
    \mathbf{S}\mathbf{v}=0
\end{equation}

This implies that at least  one eigenvalue vanishes $\lambda_{\pm}=0$,  a condition that  is satisfied when,

\begin{equation}
    r=\pm t.
\end{equation}

This absorption mechanism, known as CPA, occurs when two incident waves with an appropriate relative phase interfere destructively in the outgoing channels. For this to happen, the resonator and radiation channels need to be critically coupled, i.e. a perfect match between intrinsic and radiative losses is required $\gamma_{rad}=\gamma_{int}$.

\begin{figure}
    \centering
    \includegraphics[width=1\linewidth]{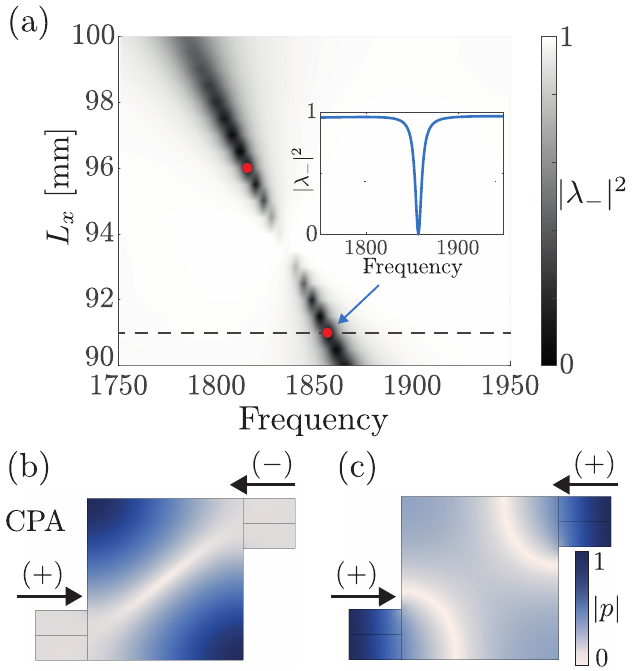}
    \caption{Realization of CPA using qBIC resonance. (a)  Eigenvalue $\vert \lambda_{-} \vert^2 $ as a function of the cavity length $L_x$ and frequency, where $\lambda_-=t-r$. The red markers denote the points where CPA is realized via qBIC formation. The inset shows the eigenvalue for the lower red point ($L_x=91$ mm). (b,c) Pressure distribution for asymmetrical (+,-) and symmetric input waves (+,+), with $L_x=91$ mm. An asymmetrical input results in CPA with strong field localization inside the cavity, whereas a symmetrical input radiates strongly in the channels. }
    \label{num}
\end{figure}

\begin{figure*}[ht]
    \centering
    \includegraphics[width=1\linewidth]{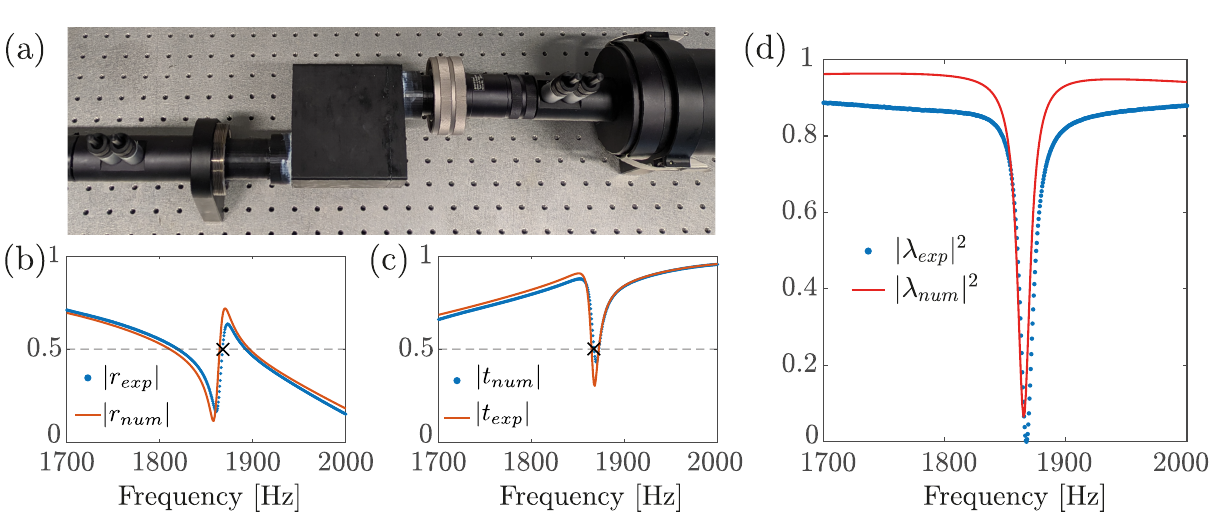}
    \caption{Experimental validation of CPA using a qBIC resonator. (a) 3D printed cavity of inner dimensions $L_x=90$mm, $L_y=94$mm and $L_z=44$ connected to two channels of inner diameter $29$mm. (b-d) Magnitude of the scattering parameters: (b) reflection coefficient $r$, (c) transmission coefficient $t$, and (d) the eigenvalue $\vert \lambda_{-}\vert^2$. The blue and red colours represent the experimental and numerical results, respectively. The cross marker corresponds to the point where $r= t$. }
    \label{FIgExp}
\end{figure*}

A straightforward implication of the critical coupling condition is that a conventional BIC (fully suppressed radiative leakage) cannot act as an absorber, since it exhibits no radiation losses ($\gamma_{ rad} = 0$). However, as discussed in the previous section, tuning the cavity length allows the  adjustment of  radiation losses, which could be employed for the realization of a high $Q$ factor absorptive filter.

In the following, we show that by carefully controlling the balance between intrinsic and radiative losses, the system can achieve CPA. Throughout this analysis, the absorption properties are discussed via the power matrix, $\Pi$,

\begin{equation}
    \Pi = \mathbf{S}^\dagger\mathbf{S},
\end{equation}

whose eigenvalues $\vert \lambda_{\pm} \vert^2$ are quantities directly related to the  dissipated power \cite{auregan1999determination}. For instance, a value of 
$1$ indicates perfect energy conservation, whereas 
$0$ corresponds to complete absorption of the incident power.

The eigenvalue of the power matrix $\vert \lambda_- \vert^2$ obtained by varying the excitation frequency and cavity length $L_x$ is shown in Fig.\ref{num}(a). The black area indicates the parameter region where the absorption is high. In particular, the eigenvalue completely vanishes for two specific lengths, $L_x \approx 96.5\,\text{mm}$ and $L_x \approx 91\,\text{mm}$.  To highlight the absorption, the eigenvalue $\vert \lambda_-\vert^2$ for $L_x = 91 \,\mathrm{mm}$ is provided in the inset of Fig.\ref{num}(a). Note that the other CPA configuration  ($L_x = 96.5 \,\mathrm{mm}$) provides similar results but with a frequency shift. At these configurations, the critical coupling condition $\gamma_{\mathrm{rad}} = \gamma_{\mathrm{int}}$ is satisfied through a reduction of radiation losses, leading to CPA with a high $Q$ factor ($Q_{\mathrm{CPA}} \approx 224$).

The configurations capable of achieving CPA via qBICs can also be deduced from the eigenfrequency analysis. In particular, the $Q$ factor can be approximated as $Q_{\mathrm{max}}/2$, since the maximum $Q$ factor (see Fig.~\ref{Figure2}.c) is primarily limited by intrinsic losses, i.e., $Q_{\mathrm{max}} \approx 1/(2\gamma_{\mathrm{int}})$. Introducing radiative losses to reach the critical coupling condition reduces the $Q$ factor by roughly a factor of two. Furthermore, the curve in Fig.~\ref{Figure2}(c) crosses $Q_{\mathrm{max}}/2$ at two cavity lengths, consistent with the scattering analysis above.

Moreover, the CPA condition also requires the input waves to match the corresponding eigenvectors. Figure \ref{num}(b–c) illustrates the pressure field at the CPA point achieved for $L_x=91$ mm under symmetric (in-phase) and asymmetric (out of phase) excitations. In our setup, CPA is realized when the input waves have a relative phase shift of $\pi$, in which case most of the energy is confined inside the cavity (see Fig. 3.b). By contrast, when a symmetric input is applied (Fig. 3.c), corresponding to the eigenvector associated with $\lambda_+$, the energy is predominantly radiated into the channels. Therefore, a symmetric input excites a radiant mode, whereas an asymmetric input results in a CPA with strong confinement inside the cavity.

 \textit{Experimental validation.—} 
In this section, the implementation of CPA using a qBIC acoustic resonator is investigated experimentally. The experimental setup is shown in Fig.~\ref{FIgExp}(a). It consists of a 3D printed cavity made of polylactic acid (PLA), with dimensions $L_x = 90\,\text{mm}$, $L_y = 94\,\text{mm}$, and $L_z = 44\,\text{mm}$, connected to two ducts of diameter $29\,\text{mm}$. The cavity walls are made sufficiently thick ($10$ mm) to minimize acoustic–structure coupling. The scattering coefficients $r$ and $t$ are measured experimentally using an impedance tube (B\&K, Type 4206).

The magnitudes of the experimentally measured (blue) and numerically computed (red) scattering coefficients, $r$ and $t$, are shown in Fig.~\ref{FIgExp}(b–c). A good agreement is observed between the experiment and the simulation, with both capturing the main resonance features. In addition, the reflection coefficient exhibits a characteristic Fano-like resonance. Figure~\ref{FIgExp}(d) shows the frequency dependence of the eigenvalue $\lvert \lambda_- \rvert^2$.  The experimental data (blue) exhibit a vanishing eigenvalue around $f_{CPA} \approx 1868$ Hz, which is associated with a high $Q$ factor of $Q_{exp} \approx 140$. The numerical results (in red) display good overall agreement but do not vanish. In fact, the numerical simulation predicts the CPA condition at a slightly larger cavity length, $L_x = 91$ mm, compared to $L_x = 90$ mm used experimentally. This small discrepancy is attributed to higher effective losses in the measurements, likely resulting from the surface roughness introduced by the additive manufacturing process, which has been shown to affect the effective viscosity \cite{ciochon2023impact}. Additional contributions may arise from slight energy losses due to wall vibrations.

Moreover, achieving CPA requires that the incident waves correspond to the appropriate eigenvectors, i.e., with the same phase and amplitude in our case. This aspect is analyzed via the measured pressure in the channels and is discussed in the Supplementary Information.

\begin{figure*}
    \centering
    \includegraphics[width=1\linewidth]{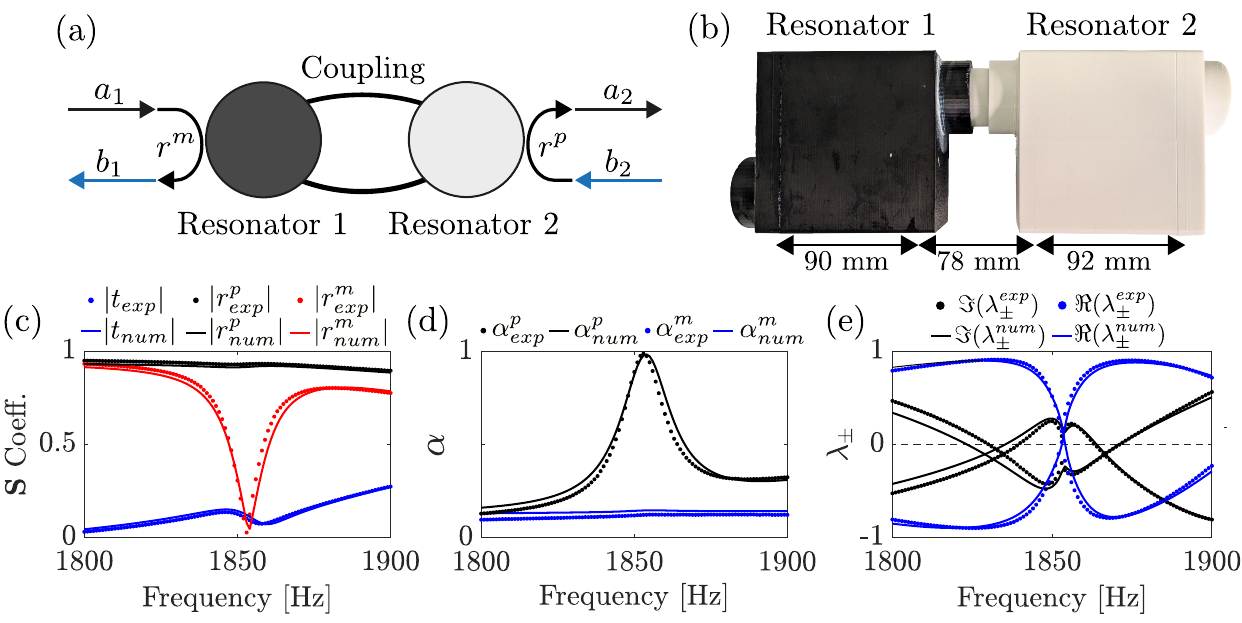}
    \caption{Emergence of an EP and unidirectional absorption using a qBIC.  (a) Sketch of the coupled system. (b) Experimental setup using two acoustic resonators. The experimental results are presented as follow: (c) Magnitude of the scattering coefficient. (d) Absorption coefficient for left $\alpha^m=1-\vert t\vert^2 -\vert r^m\vert^2$ and right impinging waves $\alpha^p=1-\vert t\vert^2 -\vert r^p\vert^2$. (e) Real and imaginary part of the eigenvalues of the scattering matrix.  }
    \label{EPs}
\end{figure*}

 \textit{Interplay between EP and qBIC.—}  
As demonstrated above, controlling the radiative losses enables to obtain a CPA. In the following, we investigate the implementation of an EP via a similar loss control approach.  To this end, we consider a new system composed of two coupled resonators, as displayed in Fig.\ref{EPs}.(a). This system can be described by the following scattering matrix and its corresponding eigenvalues,

\begin{equation}
   \mathbf{S}={\begin{pmatrix}
t & r^p \\
r^m & t
\end{pmatrix}}, \,\,\,\,\,\,\, \lambda_{\pm}=t\pm \sqrt{r^mr^p},
\end{equation}
where $r^m$ and $r^p$ represent the reflection coefficients for left and right incident waves, respectively.

Here, we focus on the formation of an EP, which occurs when both the eigenvalues $\lambda_{\pm}$ and their corresponding eigenvectors $v_{\pm}$ coalesce. This condition is satisfied when one of the reflection coefficients vanishes, yielding a degenerate pair of eigenvalues $\lambda_{\pm}=t$ and linearly dependent eigenvectors \cite{merkel2018unidirectional}. Furthermore, when this condition coincides with a vanishing transmission ($\vert t \vert=0$), both eigenvalues simultaneously vanish, resulting  in a configuration that merges absorption and EP. In fact, the system operates as a unidirectional perfect absorber, completely absorbing waves incident from the left while reflecting those incident from the right.

To implement this configuration experimentally, we employ the critically coupled resonator introduced previously, which provides CPA with a high $Q$ factor. The latter is coupled to a second cavity of identical cross-section but with a length of $92$ mm, as displayed in Fig.\ref{EPs}.(b). The second (white) resonator is engineered to act predominantly as a reflector near the CPA frequency (see Supplementary Information for its scattering coefficients). By adjusting the spacing between the two resonators, FP interference enables perfect absorption of waves incident from the left. Conversely, waves incident from the right encounter the white resonator, which reflects them almost entirely $\vert r^p\vert\approx 1$. In the experiments that follow, a duct length of
$78$ mm was selected after fine tuning (see supplementary information).

The measured scattering coefficients shown in Fig.~\ref{EPs}(c) indicate that the transmission remains below $0.3$ across the selected frequency range, while the reflection for waves incident from the right is broadband and close to unity. In contrast, for waves impinging from the left, a pronounced reflection dip appears near the CPA frequency, reaching $\lvert r^m \rvert \approx 0.025$ at $f_{EP}=1853$ Hz. At this frequency, left-incident waves are almost completely absorbed, yielding a narrowband absorption of $\alpha^m \approx 0.99$, as highlighted in Fig.~\ref{EPs}(d), whereas right-incident waves are predominantly reflected. As discussed, the unidirectionality of absorption is linked to the formation of an EP. The latter can be observed in Fig.~\ref{EPs}(e), where both real and imaginary parts of $\lambda_{\pm}$ simultaneously approach zero near $f_{EP}$. Note that the numerical results were adjusted to better match the experimental data by increasing the effective dynamic viscosity $\mu$ by a factor of 6 from its expected value at ambient conditions ($\mu_0 = 1.85 \times 10^{-5} \, \pascal  \second$). This correction accounts for the additional dissipation caused by surface roughness from the manufacturing process (FDM printing), which has been found to significantly enhance acoustic losses \cite{ciochon2024efficient}. The impact of effective viscosity in our setup is discussed in the supplementary information.

 \textit{Conclusion.—}  
 Control over wave scattering fundamentally relies on the ability to tailor dissipation, especially in applications seeking efficient absorption. In this regard, mechanisms inducing qBICs are particularly interesting as they enable precise tuning and significant suppression of radiative losses. Here, we harness qBIC physics to achieve high $Q$ absorption in airborne acoustic setups. In a two-port configuration, CPA is achieved by controlling the decay rates of a Friedrich–Wintgen resonance and tuning the system to satisfy the critical coupling condition, yielding a $Q$ factor of $140$. Afterward, two distinct resonators are coupled through an acoustic channel, where FP interference at a specific channel length induces a narrowband scattering EP. At this EP, waves incident from one side are almost completely absorbed, while waves from the opposite side are predominantly reflected, resulting in unidirectional absorption.

While this study focuses on fundamental modes and simple cavity structures, the approach can be extended to more complex geometries, higher-order modes, and alternative qBIC-generation mechanisms. The incorporation of active acoustic components, for instance via electroacoustic elements \cite{padlewski2023active}, could enable further functionalities  leveraging the presence of gain and loss units, such as CPA-lasing \cite{sakotic2023non}, which combines strong wave amplification with absorption.

\section*{Acknowledgments}
This work has been supported by l’Agence de l’Innovation de Défense (AID) in the context of the joint research laboratory MOLIERE, and by the IMPACT
project LUE “I-META,” part of the French PIA project “Lorraine Université d’Excellence” reference ANR-15- IDEX-04-LUE. 

\section*{DATA AVAILABILITY}

The data that support the findings of this study are available from the corresponding authors upon reasonable request.

\end{document}